\documentclass[
prc,%
10pt,%
final,%
notitlepage,%
oneside,%
twocolumn,%
nobibnotes,%
nofootinbib,
superscriptaddress,%
floatfix,%
showkeys,%
showpacs]%
{revtex4}
\usepackage{color}
\usepackage{amsfonts}
\usepackage{amsbsy}
\usepackage{mathrsfs}
\usepackage{graphicx}
\def\lsim{\mathrel{\rlap{
\lower4pt\hbox{\hskip-3pt$\sim$}}
    \raise1pt\hbox{$<$}}}     
\def\gsim{\mathrel{\rlap{
\lower4pt\hbox{\hskip-3pt$\sim$}}
    \raise1pt\hbox{$>$}}}     
\def\scr#1{\mbox{\scriptsize #1}}
\begin{document}
\title{Alternative Scenarios of Relativistic Heavy-Ion Collisions: \\
III. Transverse Momentum Spectra}
%
\author{Yu.B. Ivanov}\thanks{e-mail: Y.Ivanov@gsi.de}
\affiliation{Kurchatov Institute, 
Moscow RU-123182, Russia}
\begin{abstract}
Transverse-mass spectra, their inverse slopes and  
mean transverse masses
in relativistic collisions of heavy nuclei 
are analyzed in a wide range of incident energies  
 2.7 GeV  $\le \sqrt{s_{NN}}\le$ 39 GeV. 
The analysis is performed  within  the three-fluid model
employing three different equations of state (EoS's): a purely hadronic EoS,   
an EoS with the first-order phase transition  
and that with a smooth crossover transition into deconfined state.
Calculations 
show that inverse slopes and  
mean transverse masses of all the species 
(with the exception of antibaryons within the hadronic scenario) 
exhibit the step-like behavior similar to that observed for 
mesons and protons in available experimental data. 
This step-like behavior takes place for all considered EoS's 
and results from the freeze-out dynamics rather than is a signal 
of the deconfinement transition. 
A good reproduction of experimental inverse slopes and  
mean transverse masses for light species (up to proton) is achieved within all the 
considered scenarios. The freeze-out parameters are precisely the same as those 
used for reproduction of particles yields in previous papers of this series.  
This became possible because the freeze-out stage is not completely equilibrium. 
\end{abstract}
\maketitle

\section{Introduction}

This paper continues a series of reports on simulations of relativistic heavy-ion collisions within 
different scenarios \cite{Ivanov:2012bh,Ivanov:2013mxa,Ivanov:2013wha,Ivanov:2013yqa,Ivanov:2013cba}. 
These simulations were performed within a model of the three-fluid 
dynamics (3FD) \cite{3FD} employing three different equations of state (EoS): a purely hadronic EoS   
\cite{gasEOS} (hadr. EoS)
and two versions of EoS involving the deconfinement 
transition \cite{Toneev06}. These two versions are an EoS with the first-order phase transition
(2-phase EoS) 
and that with a smooth crossover transition (crossover EoS). 
Details of these calculations are described in the first paper of this series 
\cite{Ivanov:2013wha} dedicated to analysis of the baryon stopping. 
The main questions addressed in these simulations are: {\em Where and how does onset of deconfinement happen? 
What is the order of the deconfinement transition at high baryon densities?}

In this paper I report results on transverse-mass spectra, incident-energy dependence of 
inverse slopes of these spectra and mean transverse masses  
in relativistic heavy-ion collisions in the energy range from 2.7 GeV 
to 39 GeV in terms of center-of-mass energy ($\sqrt{s_{NN}}$). This domain covers 
the energy range of the beam-energy-scan program at the 
Relativistic Heavy-Ion Collider (RHIC) at Brookhaven National Laboratory (BNL)
and the low-energy-scan program at Super Proton Synchrotron (SPS)
of the European Organization for Nuclear Research (CERN), energies of newly constructed 
Facility for Antiproton and Ion Research (FAIR) in Darmstadt and the
Nuclotron-based Ion Collider Facility (NICA) in Dubna,  as well as
 the Alternating Gradient Synchrotron (AGS) at BNL.

Experimental data on transverse-mass ($m_T$) spectra of charged kaons produced in
central Au+Au \cite{Ahle:1999uy,Ahle:2000wq}
and Pb+Pb \cite{Afanasiev:2002mx,Alt:2007aa} collisions exhibit a 
peculiar dependence on the incident energy. The inverse slope
parameter 
of these spectra at
mid-rapidity increases with incident energy in the AGS energy domain 
and then saturates at SPS energies. 
The inverse slope parameter depends
on the transverse-mass interval of the exponential fit. The mean transverse mass 
provides an alternative measure of the $m_T$-spectra 
that is free of the above shortcoming.
Excitation functions of the mean transverse mass manifest a similar 
step-like behavior for charged kaons and also pions and protons \cite{Afanasiev:2002mx}. 
Such a behavior is not observed in proton-proton collisions.

In Refs. \cite{Gorenstein03,Mohanty03} this step-like behavior 
was associated with onset of the deconfinement transition. This assumption
was indirectly confirmed by the fact that microscopic transport
models--the Hadron-String Dynamics (HSD),  the Ultra-relativistic Quantum Molecular Dynamics
\cite{Bratkovskaya}
the Boltzmann-Uehling-Uhlenbeck (GiBUU) model 
\cite{Wagner}--based on hadronic degrees of freedom, failed to reproduce
the observed behavior of the kaon inverse slope \cite{Bratkovskaya,Wagner}.
Later, when partonic degrees of freedom were included in the HSD model 
\cite{Bratk09} (the Parton-Hadron-String Dynamics), the reproduction of the kaon 
inverse slopes became better. 
However, a good reproduction of all transverse-mass spectra
within the GiBUU model was achieved by inclusion of 
three-body collisions in terms of hadronic degrees of freedom  \cite{Larionov07}, 
i.e. by just enhancing the collisional interaction within the hadronic phase.

Hydrodynamic simulations of Ref. \cite{Hama04} succeeded to describe
this step-like behavior. However, in order to reproduce it these hydrodynamic
simulations required incident-energy dependence of the freeze-out
temperature which almost repeated the shape of the corresponding
kaon effective temperature. This happened even in spite of using
an EoS involving the phase transition into
quark-gluon plasma (QGP). This way, the problem of kaon effective
temperatures was just translated into a problem of freeze-out
temperatures. Moreover, results of Ref. \cite{Hama04} imply that
peculiar  incident-energy dependence of the kaon effective
temperature may be associated with dynamics of freeze-out.

In Refs. \cite{3FDpt,3FDpt-long} it was shown that dynamical description of the
freeze-out \cite{Russkikh:2006aa,Ivanov:2008zi}, accepted in the 3FD model, naturally explains the step-like 
behavior of inverse-slope parameters even without any deconfinement transition in the EoS. 
This freeze-out dynamics, effectively resulting in a
pattern similar to that of the dynamic liquid--gas transition, differs
from conventionally used  freeze-out schemes.
This explanation is equally applicable to the energy dependence of the mean transverse mass. 
Later, within the Hybrid Hydro-Kinetic Model \cite{Petersen:2009mz}
it was confirmed  that different freeze-out procedures have almost
as much influence on the mean transverse mass excitation function as the EoS.

In this paper I demonstrate that inverse slopes and mean transverse masses 
of various species 
(with the exception of antibaryons within the hadronic scenario)
exhibit step-like behavior for all considered EoS's. This is 
only the effect of the dynamical freeze-out \cite{Russkikh:2006aa,Ivanov:2008zi} accepted in the model. 
Moreover, the inverse slopes, mean transverse masses and particle yields are 
described within precisely the same freeze-out procedure, 
contrary to the common belief that the kinetic freeze-out 
should happen later that the chemical one. 
The reason why the unique freeze-out works both for kinetic and 
chemical quantities is described in the next section (sect. \ref{Model}).

As was demonstrated in the first papers of this series 
\cite{Ivanov:2012bh,Ivanov:2013wha,Ivanov:2013yqa}, 
onset of the  deconfinement transition takes place in the region of top-AGS--low-SPS 
incident energies within the considered here first-order-transition
and crossover scenarios. The experimental baryon stopping indicates 
certain signs of a  deconfinement transition \cite{Ivanov:2012bh,Ivanov:2013wha}
in this energy region. 
The hadronic scenario fails to reproduce antibaryon production \cite{Ivanov:2012bh,Ivanov:2013wha,Ivanov:2013yqa} 
above this energy region, while the  deconfinement scenarios do.  
The change of behavior of experimentally available excitation functions of inverse slopes and 
mean transverse masses also occurs in this energy range \cite{Gazdzicki:2010iv}. 
Therefore, in this paper the attention is 
primarily focused on this 
incident energy range. Before proceeding to discussion of 
inverse slopes and mean transverse masses it is reasonable to consider 
the transverse mass spectra themselves, which are the source data
for the former ones.

\section{3FD Model}\label{Model}

A conventional way of applying the fluid dynamics to heavy-ion collisions at
RHIC and LHC energies starts from an initial state 
that is prepared by 
means of various kinetic codes \cite{Bleicher08,Bleicher09,Hama:2004rr,Nonaka:2012qw}.
Such approaches 
disregard effects of a possible deconfinement transition at the stage of inter-penetration of 
colliding nuclei, and hence cannot be used for searching signals of deconfinement
at this stage. 
Contrary to these approaches, the 3FD model treats the collision process from the 
very beginning, i.e. from the stage of cold nuclei up to  freeze-out, within  
the fluid dynamics. 

In order to take into account a finite stopping
power at the initial stage of the nuclear collision, 
the 3FD model deals with two baryon-rich fluids which simulate 
a counter-streaming regime of leading
baryon-rich matter initially associated with constituent nucleons of the projectile
(p) and target (t) nuclei.   
In addition, newly produced particles,
populating the mid-rapidity region, are associated with a fireball
(f) fluid.
Therefore, the 3-fluid approximation is a minimal way to 
simulate the finite stopping power at high incident energies.
Each of these fluids is governed by conventional hydrodynamic equations
which contain interaction terms in their right-hand sides. 
These interaction terms describe mutual friction of the fluids and 
production of the fireball fluid.
In terms of the above-mentioned conventional applications of the one-fluid hydrodynamics, 
the friction results in production of an initial state for the fluid evolution, i.e. 
it gives rise to the intial equilibration of the colliding matter.

The friction between fluids was fitted to reproduce
the baryon stopping observed in (net)proton rapidity distributions for each EoS, 
as it is described in  Ref. \cite{Ivanov:2013wha} in detail.
The baryon stopping turns out to be only 
moderately sensitive to the freeze-out energy density. 
The freeze-out energy density $\varepsilon_{\scr frz}=$ 0.4 GeV/fm$^3$ 
was chosen mostly on the condition of the best reproduction 
of secondary particle yields.

\begin{figure*}[tbh]
\includegraphics[width=17.0cm]{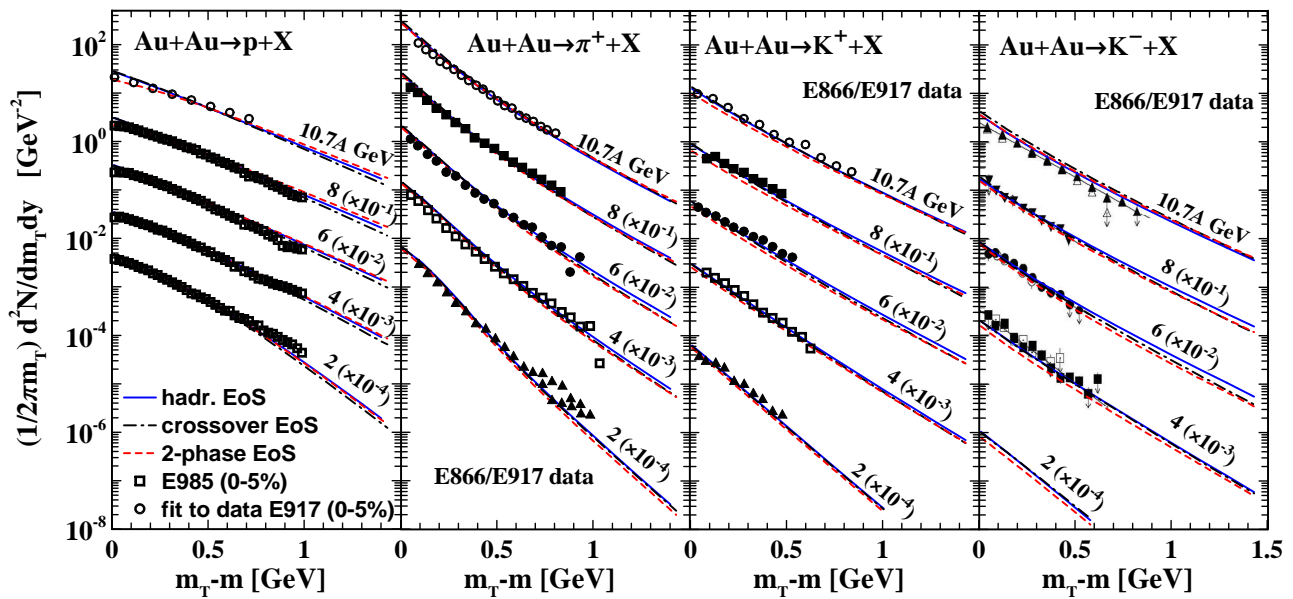}
 \caption{(Color online)
Transverse mass spectra (at midrapidity) of 
protons (data from E917 \cite{Back:2002ic} at 10.8$A$ GeV and E895 \cite{Klay:2001tf} at 2-8$A$ GeV), 
positive pions (data from E866 and E917 \cite{Ahle:1999uy}), and positive (E866 and E917 \cite{Ahle:1999uy}) 
and negative (E866 and E917 \cite{Ahle:2000wq}) kaons
from central collisions  Au+Au (5\% centrality) at AGS incident energies, 
$E_{\rm lab}=$ 2$A$, 4$A$, 6$A$, 8$A$ and 10.7$A$ GeV, calculated with different EoS's 
at impact parameter $b=$ 2 fm.   
} 
\label{fig_ags-mt}
\end{figure*}
\begin{figure}[htb]
\includegraphics[width=5.0cm]{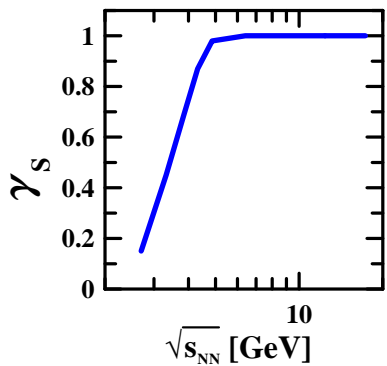}
 \caption{(Color online)
Strangeness suppression factor for central Au+Au collisions as a function of the center-of-mass energy
 of colliding nuclei. 
}  
\label{fig-gammaS}
\end{figure}
\begin{figure*}[p]
\includegraphics[width=15.2cm]{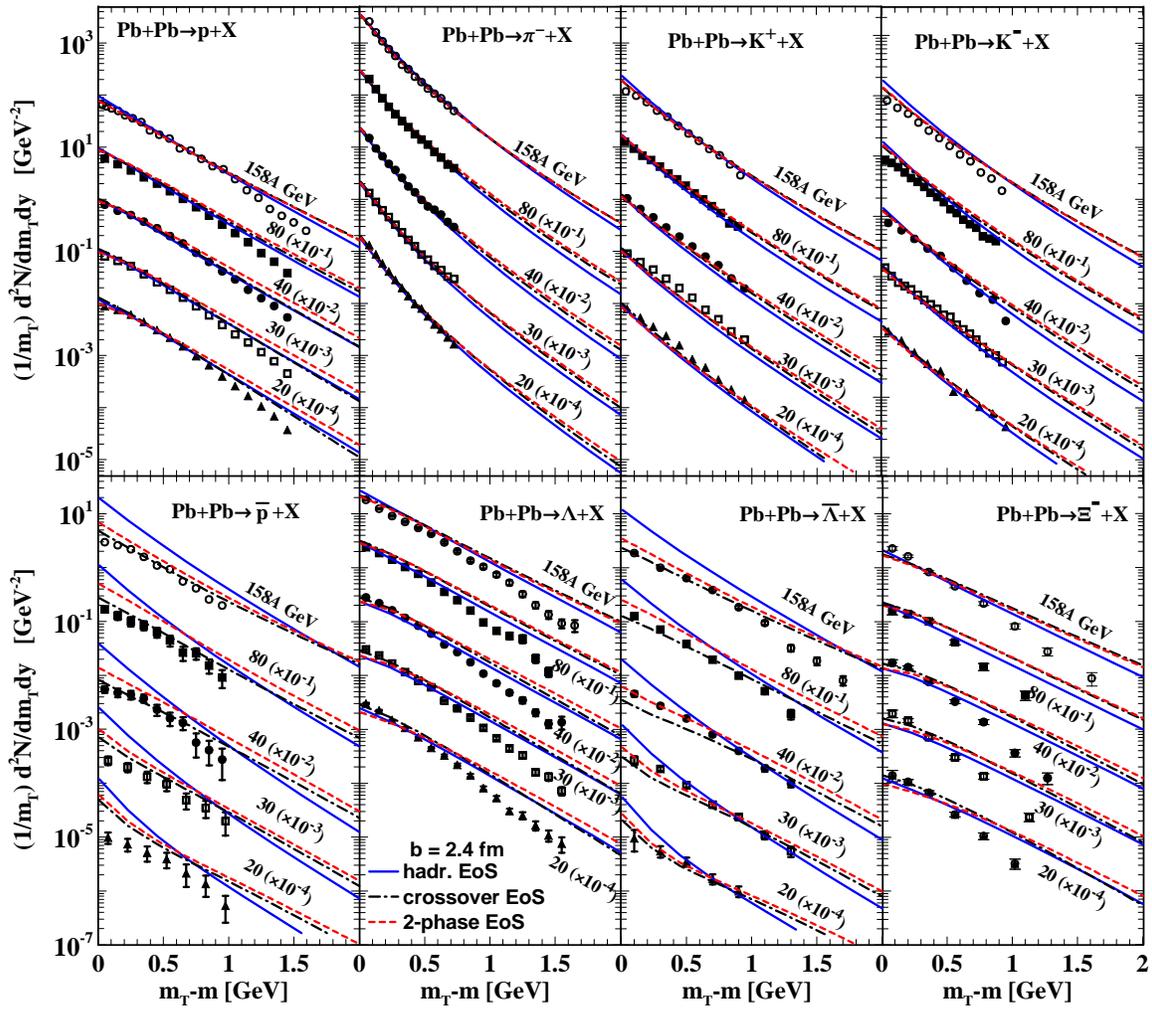}
 \caption{(Color online)
Transverse mass spectra (at midrapidity) of 
protons,  pions,    kaons, antiprotons, 
$\Lambda$, $\bar{\Lambda}$ and  $\Xi^-$ hyperons 
from central Pb+Pb collisions ($b=$ 2.4 fm) at SPS incident energies.   
Experimental data are from 
NA49 Collaboration
\cite{Gazdzicki:2004ef,Afanasiev:2002mx,Alt:2007aa,Alt:2006dk,Anticic:2004yj,Anticic:2003ux,Alt:2008qm}. 
Calculations for $\Lambda$, $\bar{\Lambda}$ and $\Xi^-$ hyperons at $E_{lab}=$ 158$A$ GeV were performed 
at $b=$ 4.6 fm because the centrality selection of the respective data \cite{Alt:2008qm} is  
0-10\%, contrary to other presented data corresponding to either 0-5\% or 0-7\% selection. 
} 
\label{fig_sps-mt}
\end{figure*}
\begin{figure*}[thb]
\includegraphics[width=17.0cm]{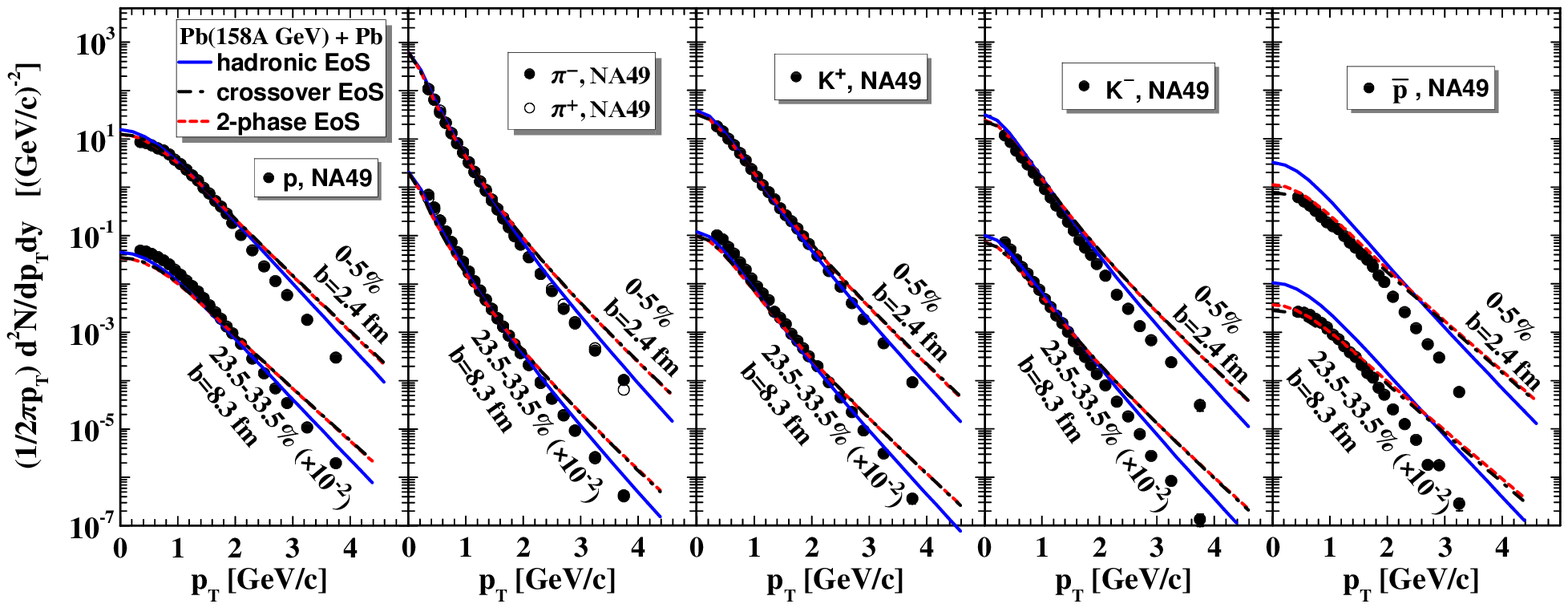}
 \caption{(Color online)
Transverse momentum spectra (at midrapidity) of 
protons, pions, kaons, and antiprotons 
from central collisions  Pb+Pb ($b=$ 2.4 fm) at 158$A$ GeV in the extended $p_T$ range.   
Experimental data are from NA49 Collaboration
\cite{Alt:2007cd}. 
} 
\label{fig_sps-high-pt}
\end{figure*}

It is important that the same freeze-out is used to describe both the chemical (particle abundances) 
and kinetic (energy and momentum spectra) observables. On the other hand, it is commonly 
accepted that chemical observables require higher freeze-out temperatures than the 
kinetic ones, that implies that the kinetic freeze-out occurs somewhat later that the 
chemical one. This is indeed true, if the system is assumed to be completely thermodynamically 
equilibrium at the freeze-out stage. However, it is not the case in the 3FD model. 

In the 3FD model  
 the baryon-rich fluids are either unified (i.e. mutually 
stopped and merged into a single unified fluid) or spatially separated 
to the instant of the freeze-out. However, the 
baryon-free f-fluid still keeps its identity. During the whole collision process, the  
interaction between the f-fluid and the baryon-rich ones causes their unification. 
Nevertheless, till the very freeze-out this unification is not complete. 
Therefore, the freezing-out system consists of two overlapping fluids and hence, strictly speaking,  
is not thermodynamically equilibrium. 
Then, the temperatures occur to be low enough to reproduce the kinetic observables, 
while  abundances of the secondary particles are enhanced by a contribution of the 
f-fluid to the extent that they also agree with data. 
If the unified baryon-rich fluid and the f-fluid are artificially unified into 
a single fluid, as it was done in Ref. \cite{Ivanov:2013cba}, 
then the temperature and baryon chemical potential of this unified fluid
well reproduce (within deconfinement-transition scenarios) 
the corresponding freeze-out parameters deduced from experimental data 
within the statistical model \cite{Andronic:2008gu}.

In Ref. \cite{Satarov:2012yk}
a comparison of results of transport (GiBUU) and hydrodynamic
calculations for the expansion of a baryon–rich hadronic fireball.
Strong deviations from chemical equilibrium, especially at the final (freeze-out) stage, 
were found within the transport calculations. This nonequilibrium results in 
an enhancement of yields of the secondary particles. 
Within the 3FD model the contribution of the f-fluid at the freeze-out precisely simulates this 
nonequilibrium enhancement.

The 3FD model \cite{3FD} is a
straightforward extension of the 2-fluid model with radiation of
direct pions \cite{MRS88,gsi94,MRS91} and (2+1)-fluid model
\cite{Kat93,Brac97}. The above models were extend in such a
way that the created baryon-free fluid (which is called a
``fireball'' fluid, following the Frankfurt group) is treated
on equal footing with the baryon-rich ones. In addition, 
a certain formation time $\tau$ is attributed to the fireball fluid, during
which the matter of the fluid propagates without interactions. 
The formation time 
is  associated with a finite time of string formation. It is similarly 
incorporated in kinetic transport models such as 
Ultrarelativistic Quantum Molecular Dynamics (UrQMD) \cite{Bass98} 
and HSD \cite{Cassing99}.


\section{Transverse Mass Spectra}
\label{Transverse Spectra}

In this section I report results on transverse-mass ($m_T$) spectra of various species from central 
Au+Au (for AGS energies) 
and Pb+Pb (for SPS energies) collisions. 
Correspondence between experimental centrality, i.e the fraction of the total reaction 
cross section related to a data set,  
and the mean value of  the impact parameter ($b=$ 2.4 fm for centra Pb+Pb collisions) 
is taken from the paper \cite{Alt:2003ab}
in case of NA49 data. 
For central Au+Au collisions the value of $b=$ 2 fm is approximately estimated proceeding from geometrical considerations. 
A contribution of weak decays of strange hyperons into non-strange hadron yields is disregarded
in accordance with measurement conditions of the NA49 collaboration. 
At the AGS energies the contribution of weak decays is negligible. 
3FD calculations for RHIC energies are not presented because   
the RHIC data for identified particles in the considered energy range
are very fragmentary and/or have a preliminarily status 
\cite{Zhu:2012ph,Abelev:2009bw}. Therefore, 
predictions for the RHIC energies 
are done in terms of inverse slopes and mean transverse masses, see Sect. \ref{Inverse Slopes}. 
At RHIC energies, contributions of weak decays of strange hyperons 
into non-strange hadron yields were included 
in accordance with measurement conditions of the STAR and PHENIX  collaborations.

The $m_T$ spectra at mid-rapidity from central collisions at AGS energies 
are presented in Fig. \ref{fig_ags-mt}. 
As seen, the experimental data are reasonably good reproduced within all considered 
scenarios. All these scenarios predict approximately the same results for 
all displayed species 
except for protons at 10$A$ GeV within the 
2-phase EoS. 
The 2-phase scenario better reproduces the shape of the proton spectrum at 10$A$ GeV, 
though somewhat underestimates its overall normalization which 
has been already discussed in Refs. \cite{Ivanov:2012bh,Ivanov:2013wha}. 
The strangeness production at low incident
energies is overestimated within the 3FD model, since used EoS's 
are based on the grand canonical ensemble. 
Therefore, the spectra   
of single-strange particles, 
like $K^\pm$, displayed in Fig. \ref{fig_ags-mt}, 
are multiplied by $\gamma_S$ factor, that takes into 
account  an additional strangeness 
suppression due to constraints of canonical ensemble \cite{Koch:1986ud}.
The excitation function of the $\gamma_S$ factor is presented in 
Fig. \ref{fig-gammaS}, which is of course applicable only to central 
collisions of considered nuclei. 
As seen, at $E_{\scr{lab}}>$ 10$A$ GeV there is no
need for additional strangeness suppression.

\begin{figure*}[htb]
\vspace*{-3mm}
\includegraphics[width=14.0cm]{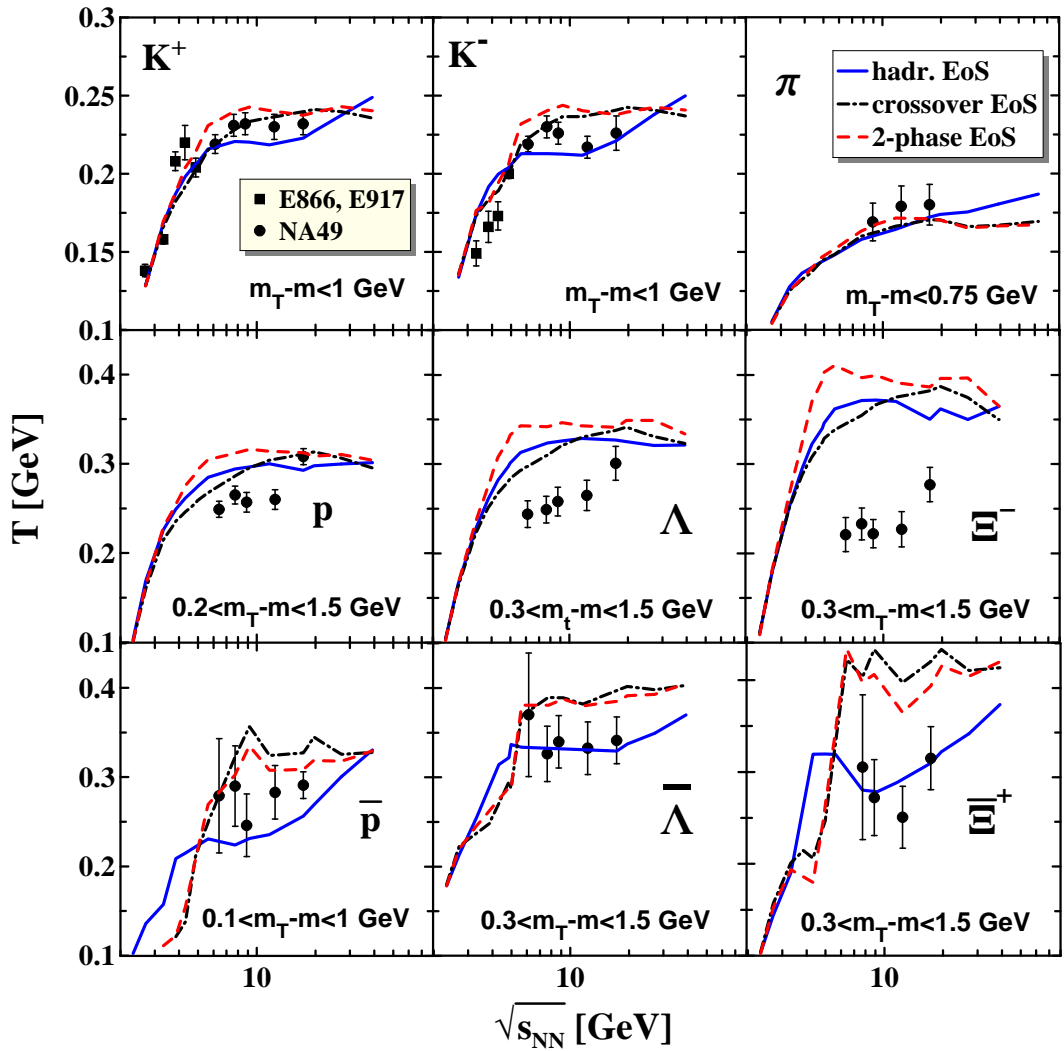}
 \caption{(Color online)
Inverse slopes of transverse-mass spectra (at midrapidity) of various species
from central collisions Au+Au (at AGS and RHIC energies, $b=$ 2 fm) and Pb+Pb  (at SPS energies, $b=$ 2.4 fm)
as functions of the center-of-mass incident energy.   
Experimental data are taken from Refs. \cite{Alt:2007aa,Alt:2006dk,Alt:2008qm}. 
The $m_T$ range, in which the exponential fit (\ref{Ttr}) was performed, is indicated
in each panel. 
} 
\label{fig_pt_slope}
\end{figure*}

The $m_T$ spectra at mid-rapidity from central collisions at SPS energies 
are presented in Fig. \ref{fig_sps-mt}. 
As seen, for abundant probes (upper raw of panels) 
agreement with data is certainly better than 
that for rare ones (lower raw of panels). This is again an artifact of the description based 
based on grand-canonical statistics which requires 
``large'' multiplicities to be valid. A lack of exact conservations (of baryon 
number and strangeness) in the grand-canonical ensemble 
results in overestimation of data for rare probes. 
The hadronic scenario fails to reproduce the antibaryon (antiproton and anti-lambda) spectra even 
at low $m_T$, as it has already been discussed in Ref. \cite{Ivanov:2013yqa}.
The 3FD predictions essentially overestimate the 
high-$m_T$  ends of these data.
This is even better seen in comparison with the NA49 data  \cite{Alt:2007cd}
taken in a wide $p_T$ range at 158$A$ GeV, see   Fig. \ref{fig_sps-high-pt}. 
This overestimation is a manifestation of finiteness 
of the considered system. Even abundant hadronic probes become rare 
at high momenta. Therefore, their treatment on the basis of grand canonical ensemble
results in overestimation of their yields. 
Moreover, the more rare probe is the hadron by itself, the stronger  its 
high-$p_T$  end of the spectrum  is suppressed due to restrictions of the canonical ensemble. 
In fact, the hadronic scenario closer reproduces the high-$m_T$ and -$p_T$  ends of the spectra
as compared with the deconfinement-transition ones. 
However, this cannot be considered as an advantage of the hadronic scenario because the 
hydrodynamics is primarily expected to describe the soft parts of the spectra.

\section{Inverse Slopes of $m_T$ Spectra and Mean Transverse Masses}
\label{Inverse Slopes}


In order to quantify the spectral
shape, the invariant $m_t$ spectra are usually fitted by an exponential function
\begin{eqnarray}
\label{Ttr}
\frac{d^2 N}{m_T \; d m_T \; d y} \propto
\exp \left(-\frac{m_T}{T(y)}  \right),
\end{eqnarray}
where $m_T$ and $y$ are the transverse mass and rapidity,
respectively, and $T(y)$ is the inverse slope parameter that generally 
depends on the rapidity. Below 
we consider only slopes at midrapidity. Therefore, the argument $y$ 
is omitted. Incident energy dependence of inverse slope parameters at midrapidity
for various species is shown in Fig. \ref{fig_pt_slope}.

The exponential function results in a very good fit for kaons at $m_T-m<$ 1 GeV. For higher $m_T$ this fit 
underestimate the data, as it was demonstrated in Ref. \cite{Ivanov:2013yqa} based on 
high-$p_T$  ends of these data \cite{Alt:2007cd}. 
Heavier particles exhibit deviation from an exponential behavior
already at moderate $m_T$. Therefore, the fit results 
depend on the  range in which the fit is performed. 
This range is indicated in each panel of Fig. \ref{fig_pt_slope}. 
Nevertheless, inverse slope
parameters $T$ extracted for different particles at different energies 
in a region up to moderately high $m_T$
is a useful tool to perform a spectacular comparison of different $m_T$ spectra.

\begin{figure*}[thb]
\includegraphics[width=14.0cm]{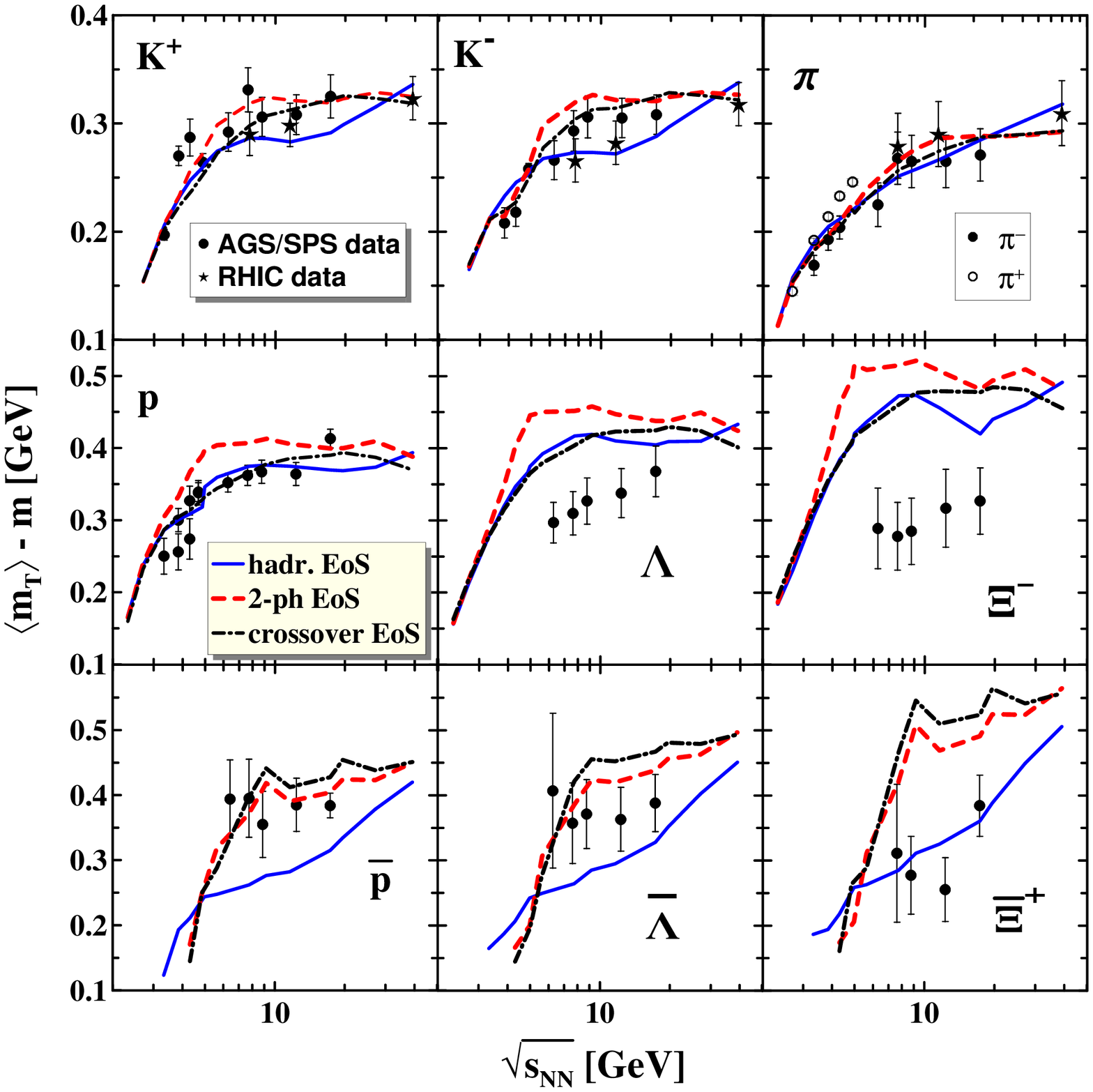}
 \caption{(Color online)
Mean transverse mass (minus particle mass) at midrapidity of various species
from central collisions Au+Au (at AGS and RHIC energies, $b=$ 2 fm) and Pb+Pb  (at SPS energies, $b=$ 2.4 fm) 
as functions of the center-of-mass  incident energy.   
Experimental data are taken from Refs. \cite{Alt:2007aa,Alt:2006dk,Alt:2008qm,Blume:2004ci}
for AGS-SPS energies and from Ref. \cite{Kumar:2011us} for RHIC energies. 
} 
\label{fig_mean-mt}
\end{figure*}
The energy
dependence of $T$ for charged kaons exhibits an interesting feature, as shown in Fig. \ref{fig_pt_slope}.
While $T$ is rapidly
rising  with center-of-mass energy for $\sqrt{s_{NN}}<$ 8 GeV, it is rather constant or only slightly
increasing above this energy. A similar observation has been made for the mean transverse mass
\begin{equation}
\label{Mtr}
\langle m_T \rangle 
= 
\frac{\displaystyle 
\int d^2 p_T \; m_T 
\left(\frac{d^2 N}{m_T \; d m_T \; d y} \right)
}{\displaystyle 
\int d^2 p_T \; 
\left(\displaystyle \frac{d^2 N}{m_T \; d m_T \; d y} \right)}
\end{equation}
of also pions and protons \cite{Alt:2007aa,Blume:2004ci}, 
i.e. of those species experimental data for which 
are available below SPS energies, see Fig. \ref{fig_mean-mt} 
The mean transverse mass is a good measure of the transverse spectrum even 
if it is not exponential like in (\ref{Ttr}). 
In proton-proton collisions such a behavior is not observed
\cite{Kliemant:2003sa}.
Therefore, the step-like behavior of inverse slopes and mean transverse masses 
certainly results  from collective motion of the matter.

Calculations within the 3FD model show that inverse slopes and mean transverse masses
of all the species exhibit the 
same step-like behavior for all considered EoS's. The exception is the mean $\langle m_T \rangle$
of antibaryons within hadronic scenario, which fails to reproduce any 
antibaryon observables \cite{Ivanov:2013yqa}. 

\begin{figure}[thb]
\includegraphics[width=7.0cm]{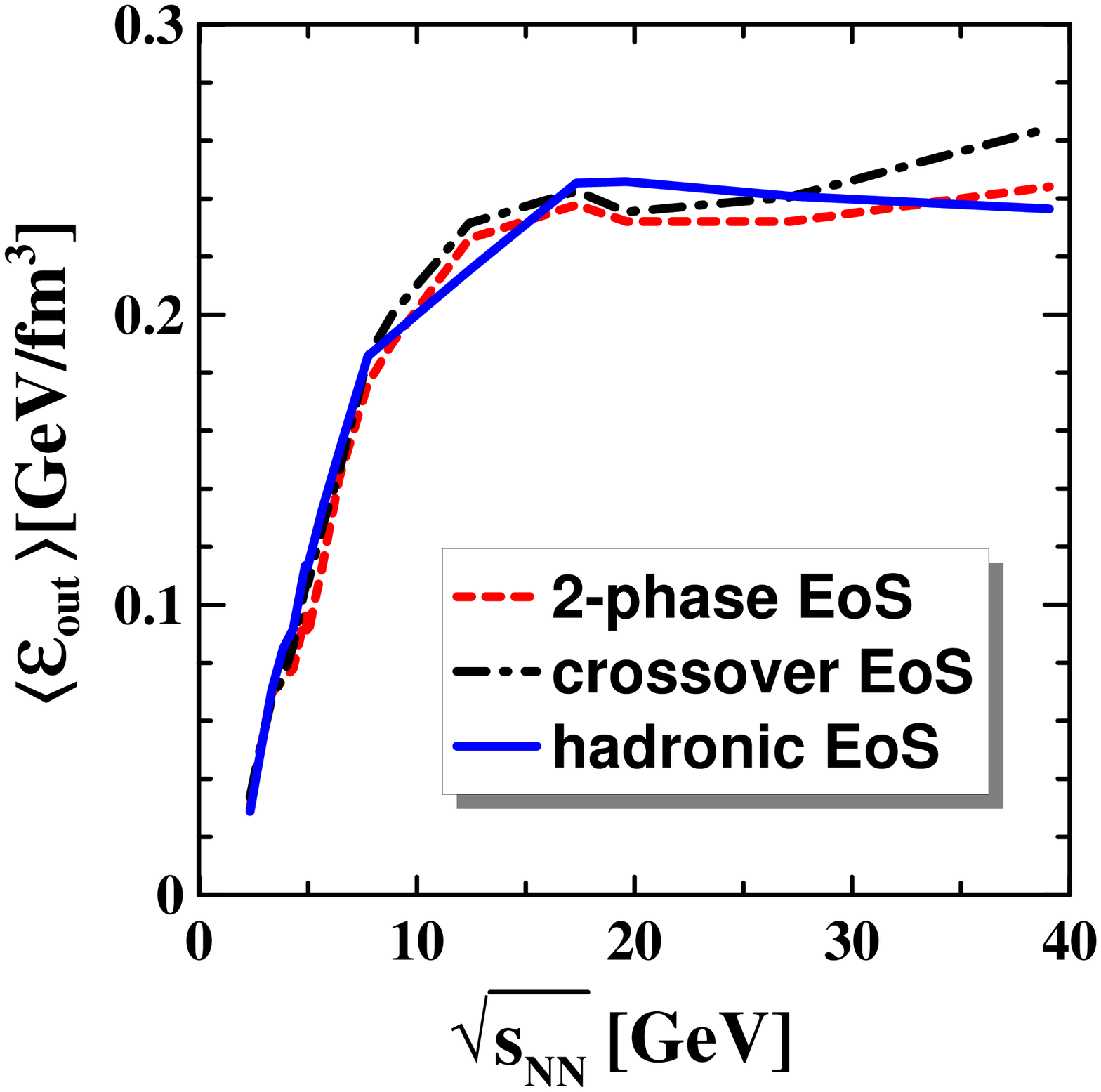}
 \caption{(Color online)
Mean energy density 
of the frozen-out matter
as a function of incident energy.   
} 
\label{fig_mean-eps}
\end{figure}

This step-like behavior is a consequence of the step-like behavior of the effective 
freeze-out energy density $\langle\varepsilon_{\scr{out}}\rangle$, see  
Fig. \ref{fig_mean-eps}. In fact, this explanation of the step-effect is similar to 
that in hydrodynamic simulations of Ref. \cite{Hama04}, where the  step-like
freeze-out temperature dependence on incident energy was required to reproduce 
the inverse-slope excitation fuctions of kaons. The 3FD model proceeds somewhat 
further as compared with Ref. \cite{Hama04} and explains the origin of this 
step-like behavior, however, not the height of the step itself that is still
related to the value of the phenomenological parameter--the freeze-out 
energy density $\varepsilon_{\scr{frz}}$. This parameter is the same for all 
EoS's and all incident energies\footnote{Only for the lowest considered incident energy
of 2$A$ GeV it was taken different: 0.3 GeV/fm$^3$}: 
$\varepsilon_{\scr{frz}}= 0.4$ GeV/fm$^3$.  
Contrary to the effective 
freeze-out energy density $\langle\varepsilon_{\scr{out}}\rangle$ at which the freeze-out 
actually happens, the $\varepsilon_{\scr{frz}}$ quantity has a meaning of 
a ``trigger'', that indicates possibility of the freeze-out.

The freeze-out procedure adopted in the 3FD model 
was analyzed  in detail in Ref. \cite{Russkikh:2006aa}. 
This method of freeze-out can be called dynamical, since the
freeze-out process here is integrated into the fluid dynamics through
hydrodynamic equations. The freeze-out
front is not defined just ``geometrically'' on the condition of the
freeze-out criterion met\footnote{The freeze-out criterion 
demands that the energy density of the matter is lower than the value of
$\varepsilon_{\scr{frz}}$.} 
but rather is a subject the fluid evolution.
It competes with the fluid flow and not always reaches the place where
the freeze-out criterion is first met.
This kind of freeze-out is similar to the model of ``continuous
emission'' proposed in Ref. \cite{Sinyukov02} and further developed in 
Refs. \cite{Sinyukov08,Knoll08}. 
There the particle emission
occurs from a surface layer of the mean-free-path width. In the 3FD model the
physical pattern is similar, only the mean free path is shrunk to zero 
(in practice, to the width of the grid cell).

The physical pattern behind this freeze-out resembles the process of 
expansion of compressed and heated classical fluid into vacuum. Physics of this 
process is studied both experimentally and theoretically
\cite{evap76,evap82,evap92,evap99,evap05}. Evaporation from free surface of normal 
(not superheated) fluid is a very slow process. Accordingly, the freeze-out of 
matter of high density ($\varepsilon > \varepsilon_{\scr{frz}}$) is suppressed in
the 3FD model. 
During expansion the fluid becomes more rarefied, still remaining quite hot. Thus, 
the fluid becomes superheated at $\varepsilon < \varepsilon_{\scr{frz}}$. 
It occurs first at the periphery of the system, 
which is first affected by the decompression wave. Evaporation from free surface 
of superheated fluid is already a fast process. Accordingly, the freeze-out 
is allowed, but not necessarily happens actually, at $\varepsilon < \varepsilon_{\scr{frz}}$.

Situations are possible, when the freeze-out criterion 
is met in the whole slab near the free surface rather than only at the surface. 
Such situations are illustrated in \cite{Russkikh:2006aa}. 
Here we have a choice either to instantaneously 
freeze out this whole near-surface slab or to wait until the freeze-out front 
will gradually traverse this slab (if ever). This choice is relied on 
results of experiments on evaporation from superheated fluids. It was shown (see, 
e.g., Ref. \cite{evap99}) that the evaporation front propagates with respect 
to fluid not faster than with the speed of sound. Precisely this choice is 
realized in the 3FD model. 
Only matter in the surface layer gets frozen out and removed from the fluid evolution
during a single time step, while inner parts of the matter 
keep on evolving hydrodynamically even in spite of meeting the the ''trigger'' freeze-out 
criterion. Thus, the matter can turn out to be over-rarefied to the instant of its 
freeze-out at the surface. 
This implies that the freeze-out front may 
stay at essentially lower energy densities than $\varepsilon_{\scr{frz}}$ 
because supersonic fluid expansion prevents it from reaching the region, where 
the condition $\varepsilon < \varepsilon_{\scr{frz}}$ is first met. 
This is precisely the case at low incident energies, 
as it was demonstrated in Ref. \cite{Russkikh:2006aa}.

Physically it implies that a particle is evaporated ("frozen-out") only if it 
escapes from the system without collisions. Thus, its mean free path 
($\lambda_{\scr{mfp}}$) should be 
larger than its path to the free surface (with due account of the future 
evolution of the fluid). Precisely this criterion is applied in the 
model of ``continuous emission'' \cite{Sinyukov02}. In the simplified 3FD version 
of the ``continuous emission'', $\lambda_{\scr{mfp}}=0$ (in practice, the cell width) 
in the fluid phase and 
$\lambda_{\scr{mfp}}\to\infty$ in the gas phase. Therefore, a particle can escape 
only from the free surface that cannot move inward the system faster than  
with the speed of sound \cite{evap99}.

The only exception from this rule is done at the final stage of the freeze-out. 
As it was observed in experiments with classical fluids  (see, 
e.g., Ref. \cite{evap82}), a fluid transforms into 
gas by explosion, if it is strongly superheated all over its volume. Therefore, 
at the final stage of the freeze-out, when criterion  is met 
in the whole volume of the fluid residue, it is assumed that the whole residue becomes 
frozen out simultaneously. In particular, this is the reason why $\varepsilon_{\scr{frz}}$
was chosen to be smaller for the lowest considered incident energy
of 2$A$ GeV: $\varepsilon_{\scr{frz}}$($E_{lab}=$ 2$A$ GeV) = 0.3 GeV/fm$^3$. 
That was done because of a large contribution of the bulk freeze-out. 
In the energy range of 4--10$A$ GeV the bulk freeze-out is not dominant and 
the freeze-out front does not reach the region, where 
$\varepsilon = \varepsilon_{\scr{frz}}$. Hence, the 3DF results in this  
energy range only weakly 
respond to variation of the freeze-out parameter $\varepsilon_{\scr{frz}}$.

Of course, the freeze-out criterion, based on the energy density,  is not universal. In particular, it is 
not applicable to the cold nuclear matter, which has $\varepsilon\approx$ 0.15 
GeV/fm$^3$ in its ground state. Therefore, the freeze-out procedure includes an additional condition, 
preserving the cold nuclear matted from being frozen out. It looks like this criterion  is 
good enough for a restricted domain of the phase diagram, where freeze-out 
of hot nuclear matter really occurs.

The "step-like" behavior of $\langle\varepsilon_{\scr{out}}\rangle$ (see  
Fig. \ref{fig_mean-eps}) 
is a consequence of the freeze-out dynamics, as it was demonstrated in Ref.
\cite{Russkikh:2006aa}. At low (AGS) incident energies, the energy density
achieved at the border with vacuum, $\varepsilon^s$, is lower than
$\varepsilon_{\scr{frz}}$. Therefore, the surface freeze-out starts
at lower energy densities. It further proceeds at lower densities up
to the global freeze-out because the freeze-out front moves not
faster than with the speed of sound, like any perturbation in the
hydrodynamics. Hence it cannot overcome the
supersonic barrier and reach dense regions inside the expanding system.
With the incident energy rise the energy density achieved at the
border with vacuum gradually reaches the value of
$\varepsilon_{\scr{frz}}$ and then even overshoot it. If the
overshoot happens, the system first expands without freeze-out. The
freeze-out starts only when $\varepsilon^s$ drops to the value of
$\varepsilon_{\scr{frz}}$. Then the surface freeze-out occurs really
at the value $\varepsilon^s \approx \varepsilon_{\scr{frz}}$ and
thus the actual freeze-out energy density saturates at the value
$\langle\varepsilon_{\scr{out}}\rangle \approx
\varepsilon_{\scr{frz}}/2$, i.e. at the half fall from $\varepsilon^s$ to zero. 
This freeze-out dynamics is quite stable
with respect to numerics \cite{Russkikh:2006aa}.

It is convenient to discuss inverse slope parameters $T$ in terms of collective properties 
of the frozen out matter, i.e. in terms of 
the freeze-out temperature $T_{\rm frz}$ and transverse velocity $v_{\rm tr}$. 
At moderate $v_{\rm tr}$, the relation between  $T$ and ($T_{\rm frz},v_{\rm tr}$)  
 approximately reads 
\begin{eqnarray}
\label{m-dep}
T \approx T_{\rm frz} + \frac{1}{2} m v_{\rm tr}^2, 
\end{eqnarray}
where $m$ is the particle mass. 
This relation results from the nonrelativistic limit of the  
blast-wave model \cite{Bondorf78,Siemens79,Schnedermann:1993ws}. 
The  freeze-out temperature $T_{\rm frz}$ and transverse velocity $v_{\rm tr}$ are 
assumed to be the same for all species, as they are collective quantities of the matter. 
Notice that the mean transverse mass is identical to the inverse slope parameter, it the 
spectrum is precisely of the exponential form (\ref{Ttr}). 
Therefore, all the reasoning below is equally applicable to the mean transverse masses.

\begin{figure}[thb]
\includegraphics[width=8.5cm]{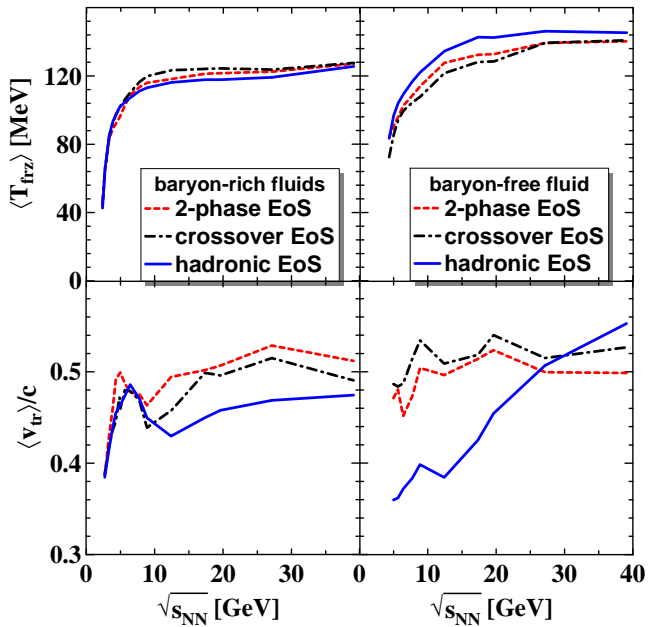}
 \caption{(Color online)
Mean temperatures ($\langle T_{frz}\rangle$) 
and  transverse velocities ($\langle v_{tr}\rangle$)  
of the frozen-out baryon-rich  (left panels) and  baryon-free fluids (right panels)
averaged over the central layer of colliding system 
(-2 fm $\le z \le$ 2 fm with $z$ being the coordinate along the beam)  
as  functions of incident energy.   
} 
\label{fig_mean-T-v}
\end{figure}

Mean freeze-out temperatures ($\langle T_{frz}\rangle$) 
and  transverse velocities ($\langle v_{tr}\rangle$)  
of the baryon-rich and  baryon-free fluids 
averaged over the frozen-out system are presented in Fig. \ref{fig_mean-T-v}
as  functions of incident energy. 
The baryon-rich fluids
are either spatially separated or 
unified at the freeze-out stage.  
The baryon-free (``fireball'') fluid
remains undissolved in baryonic fluids till the freeze-out. 
As mentioned above, the fireball fluid is characterized by  
a certain formation time $\tau$, during
which the matter of the fluid propagates without interactions. 
The main difference concerning this baryon-free fluid in considered alternative 
scenarios consists in different formation times: $\tau = 2 \;\mbox{fm/c}$ 
for the hadronic scenario and $\tau = 0.17 \;\mbox{fm/c}$ for scenarios involving 
the deconfinement transition  \cite{Ivanov:2013wha}.

As seen from simulations, the main contribution to baryon and meson yields 
comes from baryon-rich fluids. Only at highest considered energy of 
$\sqrt{s_{NN}}=$ 39 GeV approximately half of pions at the mid-rapidity are produced from the 
baryon-free fluid within the deconfinement-transition scenarios. 
For all other particles and considered energies this 
fraction is essentially lower. 
At the same time the fraction of half for pions  at the mid-rapidity from the 
baryon-free fluid is achieved already at $\sqrt{s_{NN}}\simeq$ 9 GeV 
(i.e. $E_{lab}\simeq$ 40$A$ GeV) within the hadronic scenario. 
This is one of the reasons why $\tau$ was chosen so large in the 
hadronic scenario. 
Large formation time prevents absorption of the baryon-free matter by 
the baryon-rich fluids. 
Without this large contribution of the baryon-free fluid
it is impossible to reproduce mesonic yields at SPS energies. 
However, this strongly developed baryon-free fluid makes bad job for 
antibaryons in the case of hadronic EoS. The reason is that 
antibaryons are dominantly 
produced from the baryon-free fluid even at lower considered incident energies. 
Their yields in the hadronic scenario strongly overestimate experimental 
data \cite{Ivanov:2013yqa}.

The qualitative difference of excitation functions 
of  mean transverse masses of antibaryons 
(lower panels in Fig.  \ref{fig_mean-mt})
within the hadronic scenario
from those in the deconfinement-transition scenarios 
results from difference in dynamics of the baryon-free fluid. 
To a minor degree this qualitative difference also concerns the  inverse slopes  
of antibaryons (lower panels in Fig. \ref{fig_pt_slope}). 
To illustrate this in terms of quantities of the qualitative formula
(\ref{m-dep}) let us consider mean temperatures ($\langle T_{frz}\rangle$) 
and  transverse velocities ($\langle v_{tr}\rangle$)  
of the frozen-out baryon-rich  and  baryon-free fluids 
averaged over the central region of the colliding system, see Fig. \ref{fig_mean-T-v}.
The central region is chosen because the mid-rapidity region, in particular 
mid-rapidity transverse spectra, are predominantly populated by particles from 
the central spatial region. This central spatial region is defined as a layer 
orthogonal to the beam direction ($z$) placed around origin of the $z$-axis, 
i.e. -2 fm $\le z \le$ 2 fm.

As seen from Fig. \ref{fig_mean-T-v}, 
the mean freeze-out temperatures ($\langle T_{frz}\rangle$) 
of the baryon fluids exhibit a step-like behavior because of peculiarities of the 
freeze-out process discussed above. 
The temperatures only slightly rise above the ``step'' threshold. 
The transverse velocities ($\langle v_{tr}\rangle$)  
of the baryon-rich fluids within all the considered scenarios exhibit
very similar behavior. 
A dip in the range near $\sqrt{s_{NN}}\approx$ 10 GeV 
is present in both hadronic and deconfinement-transition scenarios 
and is associated with transition from a single baryon-rich fireball at lower 
incident energies to two spatially separated fireballs at high energies. 
This is a point, where incomplete baryon stopping results in spatial 
separation of the projectile-like and target-like leading particles at the 
late stage of the evolution. It was illustrated in Ref.  \cite{Russkikh:2006aa}
for the hadronic scenario. For the deconfinement-transition scenarios
the picture is similar. 
Therefore, as it results from Eq. (\ref{m-dep}), the inverse slopes and mean 
transverse masses of all the species, except for antibaryons, exhibit the step-like behavior
with the above dip stronger or weaker revealed for different species.

The transverse velocities of the baryon-free fluid exhibit very different 
behavior for hadronic and deconfinement-transition scenarios. Whereas 
$\langle v_{tr}\rangle$ reveal saturation (up to numerical fluctuations) 
for deconfinement-transition scenarios, within the hadronic scenario 
the transverse velocity gradually grows with the incident energy rise. 
The formation time is short within 
deconfinement-transition scenarios. 
Hence, the baryon-free fluid is strongly coupled with the baryon-rich fluids 
almost from the instant of its birth.  Therefore, its evolution is very similar 
to that of the baryon fluids. Based again on Eq. (\ref{m-dep}), we can expect 
the step-like behavior of the inverse slopes and mean 
transverse masses of antibaryons, as it is the case in 
Figs. \ref{fig_pt_slope} and \ref{fig_mean-mt}, even  numerical fluctuations 
are very similar.

The situation differs if the formation time is comparatively large 
like in the hadronic scenario.  Therefore, the baryon-free fluid 
is less dragged by the baryon fluids and hence its transverse velocities
turn out to be much less correlated with those of the baryon fluids.
The baryon-free fluid evolves almost independently of the baryon-rich fluids. 
Nevertheless, a weak remnant of the dip associated 
with transition from a single baryon-rich fireball at lower 
incident energies to two spatially separated fireballs at high energies 
is seen even in this case. For deconfinement-transition scenarios, this 
weak dip cannot be well distinguished against the background of the 
above mentioned numerical fluctuations. 
The increase of the transverse velocities with the incident energy rise results from 
growing density of the baryon-free fluid 
with the incident-energy rise.
This behavior of the transverse velocities 
results in violation of the step-like behavior 
of  mean transverse masses 
(and, to minor extent, of the inverse slopes) 
of antibaryons within the hadronic scenario. 
Indeed, the second term in Eq. (\ref{m-dep}) increases with the energy especially 
for large masses $m$ which is the case for antibaryons. 
Even inverse slopes and mean 
transverse masses of light mesons start to rise at high incident energies 
because of a large contributions of the baryon-free fluid to their yields.

The inverse slope parameters deduced from experiment \cite{Blume:2011sb} 
increase linearly with particle mass up to the mass of 
$\Lambda$ and $\bar{\Lambda}$ hyperons,  
see Fig. \ref{fig_pt_slope}. Though, already for the $\Lambda$ hyperon this 
dependence is slightly violated. 
This can be understood
as a consequence of the radial expansion of the fireball, which, in a simplified picture, 
is described by Eq. (\ref{m-dep}). 
The same dependence on mass takes place for the mean transverse masses, see Fig. \ref{fig_mean-mt}.
Naturally, the same approximately linear dependence of the inverse slope on the particle mass
is predicted by 3FD calculations.

However, the data on inverse slopes of heavy hadrons beginning from the 
mass of
$\Xi^-$ and $\bar{\Xi}^+$ hyperons
do not fit into this systematics.
In Ref. \cite{Blume:2011sb} it was conjectured that 
the heavy strange particles do not participate in the radial flow to the same extent as light particles. 
The suggested in Ref. \cite{Blume:2011sb}
interpretation of this behavior is based on the assumption that rare heavy particles  
have a lower hadronic scattering cross section than light hadrons and therefore do not participate
in the radial flow that is developing during the hadronic phase of the fireball evolution. 
This leads to
the conclusion that a substantial part of the transverse expansion probed by these particles has to be
generated during the partonic phase. Thus, the rare heavy particles could be directly 
sensitive to the pressure in the early phase of the reaction. 

However, a more plausible interpretation of the violation of the linear $m$-dependence is possible. 
As it was mentioned in the previous section, rare particles are additionally suppressed 
due to restrictions of the canonical ensemble. 
Moreover, the more rare probe is the hadron by itself, the stronger  its 
high-$p_T$  end of the spectrum  is suppressed due to restrictions of the canonical ensemble. 
This high-$p_T$-enhanced suppression results in steeper slopes of the $m_T$ spectra than these 
would be in the grand canonical ensemble, i.e. in a large system. This, in its turn, 
manifests itself in a lower $T$ than it would be expected from the linear law of 
Eq. (\ref{m-dep}).

\section{Summary}
\label{Summary}

Results on transverse-mass spectra in 
relativistic heavy-ion collisions in the energy range from 2.7 GeV 
to 39 GeV in terms of center-of-mass energy, $\sqrt{s_{NN}}$, are presented.  
These simulations were performed within the 3FD model 
 \cite{3FD} employing three different EoS's: a purely hadronic EoS   
\cite{gasEOS}, and two versions of EoS involving the deconfinement 
 transition \cite{Toneev06}. These two versions are an EoS with the first-order phase transition
and that with a smooth crossover transition. 
Details of these calculations are described in the first paper of this series 
\cite{Ivanov:2013wha} dedicated to analysis of the baryon stopping.

If was found that within all scenarios the available data on $m_T$-spectra 
are reproduced approximately to the same extent 
almost for all  hadronic species (with the exception of antibaryons within the hadronic scenario) 
in the AGS-SPS energy range, i.e. from 2.7 GeV 
to 17.4 GeV in terms of $\sqrt{s_{NN}}$. 
The reproduction is better for abundant species and at low transverse masses. 
This a natural result of the fact that the model is based on grand-canonical statistics
which require high multiplicities of species to be valid. 
The grand-canonical statistics overestimates production of rare species
because it does not take onto account restrictions imposed by exact conservations 
(of strangeness, baryon charge, energy) in a finite 
system. Even abundant hadronic probes become rare 
at high momenta. Therefore, their treatment on the basis of grand canonical ensemble
results to overestimation of their yield. 
Moreover, the more rare probe is by itself, the stronger  its 
high-$m_T$  end of the spectrum  is suppressed due to restrictions of the canonical ensemble.

In the case of hadronic EoS this agreement is 
achieved at the expense of noticeable enhancement 
the inter-fluid friction in the hadronic phase \cite{Ivanov:2013wha,3FD} 
as compared with its microscopic estimate of Ref. \cite{Sat90}.
However, thus tuned hadronic scenario fails to describe $m_T$-spectra of antibaryons 
even at low transverse masses. In fact, this result was expected in view of 
the earlier reported 
failure to reproduce antibaryon rapidity distributions within the hadronic scenario
\cite{Ivanov:2013yqa}. 
The advantage of  deconfinement-transition scenarios is that they 
reproduce (with all the above-mentioned constraints) $m_T$-spectra of all species, including antibaryons,  
and do not require for that
any modification of the microscopic friction in the hadronic phase.

Excitation functions of inverse slope parameters of $m_T$-spectra of various hadrons and their 
mean transverse masses at mid-rapidity were calculated 
in the $\sqrt{s_{NN}}$ range from 2.7 GeV to 39 GeV.
Calculations within the 3FD model show that inverse slopes and  
mean transverse masses
of all the species 
(with the exception of antibaryons within the hadronic scenario) 
exhibit the 
step-like behavior similar to that observed in experimental data. 
 The exception is the mean $\langle m_T \rangle$
of antibaryons within hadronic scenario, which fails to reproduce any 
antibaryon observables \cite{Ivanov:2013yqa}.

This step-like behavior takes place for all considered EoS's and hence is not a signal 
of the deconfinement transition. 
This behavior is a consequence of the step-like behavior of the {\em effective} 
freeze-out energy density $\varepsilon_{\scr{out}}$ unlike  
the phenomenological parameter--the freeze-out 
energy density $\varepsilon_{\scr{frz}}$ which remains constant for all considered 
incident energies and has a meaning of 
a ``trigger'', which indicate possibility of the freeze-out. 
The dynamics of the freeze-out process incorporated into the 3FD model 
allows to explain how this ``trigger'' $\varepsilon_{\scr{frz}}$ value 
gives rise to the step-like behavior of the {\em effective} 
$\varepsilon_{\scr{out}}$. However, the nature of the  ``trigger'' $\varepsilon_{\scr{frz}}$ value 
still has no explanation and serves as a purely phenomenological parameter. 
In fact, similar explanation of the step-effect was indirectly implied  
in hydrodynamic simulations of Ref. \cite{Hama04}, where the  step-like
freeze-out temperature dependence on incident energy was required to reproduce 
the inverse-slope excitation fuctions of kaons. The 3FD model goes somewhat 
further as compared with Ref. \cite{Hama04} by explaining the origin of this 
step-like behavior.

Quantitative agreement with experimental data on inverse slopes  and  
mean transverse masses is achieved for ``abundant'' species, i.e. pions, 
kaons, protons, antiprotons and even anti-Lamdas. It is still surprising that  
the $\Lambda$ hyperon does not enter this list. However, data on rare probes like
$\Xi^-$ and $\bar{\Xi}^+$ hyperons turn out considerably lower than predictions 
of the 3FD model and even than data on lighter particles like 
$p$, $\bar{p}$, $\Lambda$ and $\bar{\Lambda}$. It is argued that this is 
a consequence of additional suppression due to restrictions of the canonical ensemble, 
which are not taken into account in the 3FD calculations.

All this indicates that a deconfinement-transition scenarios are certainly 
preferable in dscribing available data in the energy range from 2.7 GeV 
to 39 GeV. This conclusion agrees with those deduced in the previous papers of this 
series \cite{Ivanov:2012bh,Ivanov:2013wha,Ivanov:2013yqa}.

\vspace*{3mm} {\bf Acknowledgements} \vspace*{2mm}

I am grateful to A.S. Khvorostukhin, V.V. Skokov,  and V.D. Toneev for providing 
me with the tabulated 2-phase and crossover EoS's. 
The calculations were performed at the computer cluster of GSI (Darmstadt). 
This work was supported by The Foundation for Internet Development (Moscow)
and also partially supported  by  
 grant NS-215.2012.2.

\end{document}